\documentclass[letterpaper, 10 pt, conference]{ieeeconf} 
\IEEEoverridecommandlockouts
\usepackage{cite}
\usepackage{amsmath,amssymb,amsfonts}
\usepackage{algorithmic}
\usepackage{graphicx}
\usepackage{multirow}
\usepackage{verbatim}
\usepackage{textcomp}
\usepackage{float}  
\usepackage{subfigure}  
\usepackage{subcaption}
\usepackage{xcolor}
\usepackage{makecell}

\title{\LARGE \bf
Coordinated Energy-Trajectory Economic Model Predictive Control for Autonomous Surface Vehicles under Disturbances*
}

\author{Zhongqi Deng$^{1}$, Yuan Wang$^{1}$, Jian Huang$^{2}$, Hui Zhang$^{1}$, and Yaonan Wang$^{1}$ 
\thanks{*This work was  supported by Provincial Natural Science Foundation of Hunan Grant No. 2024JJ6159; and by National Natural Science Foundation of China Grant No. 92470202. }
\thanks{$^{1}$Zhongqi Deng, Yuan Wang, Hui Zhang, and Yaonan Wang with School of Robotics, Hunan University, Changsha, China.
        {\tt\small zhongqideng@hnu.edu.cn, yuanw@hnu.edu.cn, zhanghui1983@hnu.edu.cn, yaonan@hnu.edu.cn}}%
\thanks{$^{2}$Jian Huang with School of Artificial Intelligence and Automation, Huazhong University of Science and Technology, Wuhan, China. 
        {\tt\small huang\_jan@mail.hust.edu.cn}}%
}

\begin{document}

\maketitle
\thispagestyle{empty}
\pagestyle{empty}

\begin{abstract}
The paper proposes a novel Economic Model Predictive Control (EMPC) scheme for Autonomous Surface Vehicles (ASVs) to simultaneously address path following accuracy and energy constraints under environmental disturbances. By formulating lateral deviations as energy-equivalent penalties in the cost function, our method enables explicit trade-offs between tracking precision and energy consumption. Furthermore, a motion-dependent decomposition technique is proposed to estimate terminal energy costs based on vehicle dynamics. Compared with the existing EMPC method, simulations with real-world ocean disturbance data demonstrate the controller's energy consumption with a 0.06\% energy increase while reducing cross-track errors by up to 18.61\%. Field experiments conducted on an ASV equipped with an Intel N100 CPU in natural lake environments validate practical feasibility, achieving 0.22 m average cross-track error at nearly 1 m/s and 10 Hz control frequency. The proposed scheme provides a computationally tractable solution for ASVs operating under resource constraints.
\end{abstract}

\section{INTRODUCTION}
The autonomous surface vehicles (ASVs) have already seen some tentative applications in exploiting the streams, lakes and oceans and achieved certain successes, with a very broad prospect of application. The path following control entails tracking a predefined feasible trajectory at specified speeds and orientations \cite{7835733}. It serves as a fundamental capability for ASVs to execute mission-critical tasks. The scope and accuracy of the mission are limited by the navigation accuracy and endurance capability of the ASV. However, due to some characteristics of ASV system, including high degree of nonlinearity, parameter perturbations, multi-objective control requirements, and limited control variables, the design of controller is highly challenging \cite{9910373}. Consequently, this has led to the development of comprehensive controller designs that consider both energy consumption and accuracy.

An earlier path following study on ASV that considers the energy consumption is shown in \cite{8604920}. The authors obtain the optimal path under energy constraints through appropriate objective function based on particle swarm optimization algorithm, and further optimized the parameters of PID controller, improving the safety and economy of ASV navigation process. In addition, in response to curvilinear path following and wave interference, the authors propose an optimization method for adjusting PID parameters based on deep reinforcement learning \cite{song2022research}. Another possible ASV path following control approach is the sliding mode control in \cite{slide}, which uses a first-order surface and a second-order surface in terms of the tracking errors and different directions. 
Additionally, authors in \cite{park2017neural} propose a neural network-based output-feedback control strategy for the path following control of ASV in the presence of unknown system parameters and environmental disturbances to ensure bounded tracking errors.

As one of the most popular approaches for ASV path following control, model predictive control (MPC) is an promising  scheme to control the motion of ASV, and it has the advantage to effectively cope with various physical constraints while achieving optimal control performance \cite{10341695}. The authors in \cite{zhang} propose a robust control method that combines disturbance observer, adaptive Kalman filter, and robust model predictive control algorithm. The hybrid architecture controller address path following accuracy and rudder stability for underactuated ASVs under roll constraints. Considering the path following problem with different system parameters, in \cite{liu2021adaptive}, an adaptive model predictive control method is employed, where least squares support vector machines are utilized for controller design and online identification of different parameters.
To effectively navigate along curved paths, a new approach combined Virtual Ship Bunch and MPC is suggested in \cite{han2022tracking}, with the stability of the system is maintained through the application of Lyapunov's theorem.
In \cite{10801593}, a nonlinear MPC system is developed in wavy conditions to reduce roll during navigation.
An inspiring work in \cite{9147322} studies the energy-optimal path following control problem of the autonomous underwater vehicle in case of limited onboard
energy resources. However, the current studies that consider the cost function design are not fully account for energy consumption associated with trajectory deviations. As such, there remains potential for controller performance enhancement, particularly when addressing complex operational scenarios and large-scale mission requirements that demand higher precision.

Motivated by these considerations, the present study develops an economic model predictive control (EMPC) framework that  combines three key compoments: ship dynamics, navigation energy consumption, and path following errors. Specifically, the cross-track error is quantified as equivalent energy penalty coefficients, establishing an unified objective function that operationally links trajectory precision with energy optimization.
We  examine the energy consumption function during navigation and divide it into stage cost and terminal cost, taking into account the energy loss within and beyond the prediction horizon, thereby balancing the motion control of the unmanned vessel during path following. The simulation results with real world ocean environmental disturbance data indicate that our method not only ensures that the ASV can achieve near-optimal energy consumption but also realizes higher tracking accuracy. To validate the practical applicability of the proposed EMPC framework, field experiments are conducted in a natural lake environment. The results substantiate the method's effectiveness in the real world with time-varying environmental disturbances.

\section{ASV MODEL}

\begin{figure}[t]
    \centering
    \includegraphics[scale=0.4]{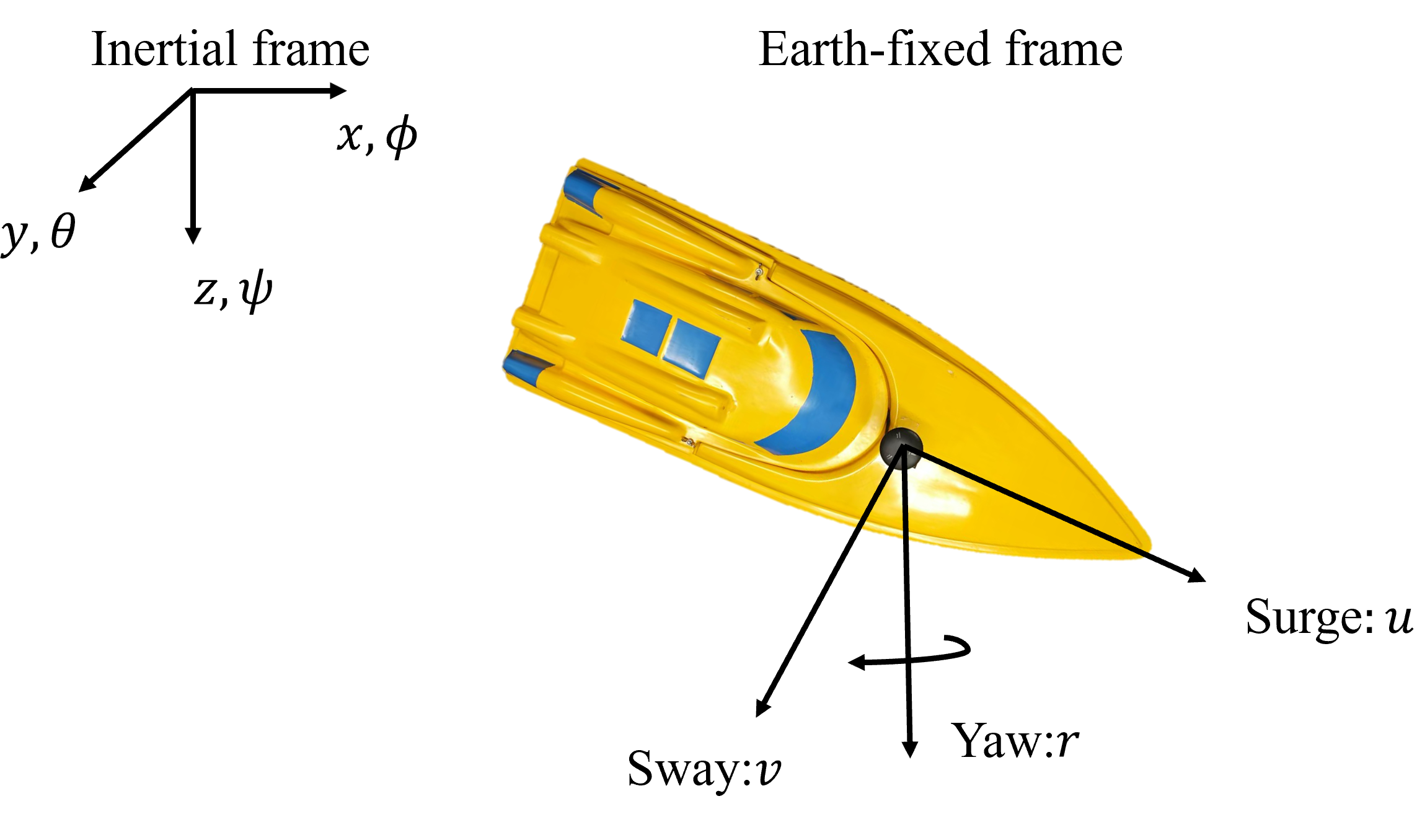}
    \caption{Reference frames and notations.}
    \label{Reference frames and notations.}
\end{figure}

To capture the vehicle motion on the water, two
reference frames are considered, as shown in Fig. \ref{Reference frames and notations.}. 
According to \cite{Fossen1994GuidanceAC} and \cite{modelnew}, the
dynamics of ASV are formulated as 
\begin{equation}
    \mathbf{M} \dot{\mathbf{v}}+\mathbf{C}(\mathbf{v}) \mathbf{v}+\mathbf{D}(\mathbf{v}) \mathbf{v}=\boldsymbol{\tau}+\boldsymbol{\tau}_d, \label{Dynamics}
\end{equation}
where $\mathbf{v}=[u\enspace v\enspace r]^T$ denotes the vehicle velocity, which contains the the vehicle surge velocity $u$, sway velocity $v$,
and yaw rate $r$ in the body-
fixed frame. The mass matrix, the skew-symmetric vehicle matrix of
Coriolis and centripetal terms, and the hydrodynamic
damping force  matrix-valued function are denoted by $ \mathbf{M}$, $ \mathbf{C(v)}$, $ \mathbf{D(v)}$ respectively. $\boldsymbol{\tau}$ is the ASV thrusts inputs vector. $\boldsymbol{\tau}_d$ is the environment disturbance.

Besides, the kinematic equation is described as
\begin{equation}
\boldsymbol{\dot{\eta}}=\mathbf{R}(\psi) \mathbf{v},   \label{Kinematics}
\end{equation}
where $\boldsymbol{\eta}=[x\enspace y\enspace \psi]^T$ is the position and orientation of the vehicle in the inertial frame. Specially, $\psi$ is the yaw angle. $\mathbf{R}(\psi)$ is a transformation matrix used to converting a state vector from body-fixed frame to inertial frame
\begin{equation}
\mathbf{R}(\psi)=\left[\begin{array}{ccc}
\cos \psi & -\sin \psi & 0 \\
\sin \psi & \cos \psi & 0 \\
0 & 0 & 1
\end{array}\right].
\nonumber
\end{equation}

Specifically, the decoupled symmetric mass matrix $\mathbf{M} \in \mathbb{R}^{3 \times 3}$ is the sum of the vehicle mass and added mass matrix
\begin{equation}
\mathbf{M}=\operatorname{diag}\left\{m_{11}, m_{22}, m_{33}\right\}.
\nonumber
\end{equation}
The matrix $\mathbf{C(v)}$ contains the rigid-body matrix and the
added mass matrix, which is denoted by 
\begin{equation}
\mathbf{C}(\mathbf{v})=\left[\begin{array}{ccc}
0 & 0 & -m_{22} v \\
0 & 0 & m_{11} u \\
m_{22} v & -m_{11} u & 0
\end{array}\right].
\nonumber
\end{equation}
Since our ASV moves at low speed, the drag
matrix $\mathbf{D(v)}$ is represented by a linear damping term 
\begin{equation}
\mathbf{D}(\mathbf{v})=\operatorname{diag}\left\{X_{u}, Y_{v}, N_{r}\right\}.
\nonumber
\end{equation}
Furthermore, the control input vector $\boldsymbol{\tau}$ relates to the ASV
thrusts as
\begin{equation}
\boldsymbol{\tau}=\left[\begin{array}{c}
T^1+T^2 \\
0 \\
\left(T^1-T^2\right) l
\end{array}\right]^T.
\nonumber
\end{equation}
Here, $T^1$ and $T^2$ are horizontal thrusters used to control the surge and yaw
motion of the vehicle, and $l$ is the distance between horizontal thrusters and the
center line.

The primary energy expenditure during vehicle operation stems from thruster propulsion. To model this consumption, we adopt the following empirical formulation for thruster power:
\begin{equation}
h_p\left(T^i\right)=\alpha\left(T^i\right)^2, \quad \text { for } i=1,2\label{h},
\end{equation}
and the energy consumption is calculated by integrating \eqref{h}.
The power conversion ratio $\alpha = 0.4364$ is derived from the data collected during thruster tests\cite{2022} .

\section{PATH FOLLOWING CONTROL PROBLEM WITH COMPREHENSIVE CONSTRAINTS}

Consider a path following mission for an ASV
with a sequence of way points, $\mathcal P = \left\{P_i \in \mathbb{R}^{2 \times 2}, i =
0,..., N\right\}$. $P_i = (x_{fi},y_{fi})$ contains the locations of the $i$ the way point. Besides, the average power required for performing a task during sailing is $T^w$.
Thus, the path following control problem under energy consumption constraint and tracking error constraint is formulated as
\begin{subequations}\label{em}
\begin{align}
 \min _{\left\{T_{k}^1\right\},\left\{T_{k}^2\right\}} &J_{E M P C}\left(\boldsymbol{X}_0,  \boldsymbol{X}_f,T_{k}^1,T_{k}^2,T^w\right\}\nonumber\\
 &=\sum_{k=0}^{H-1}J_{s}\left(T_{k}^1,T_{k}^2,T^w \right) \Delta t \nonumber\\
 &+J_{t}\left(\boldsymbol{X}_H, \boldsymbol{X}_f \right) \\
\text { s.t. } & \boldsymbol{X}_{k+1}=\mathbf{f}\left(\boldsymbol{X}_k, T_k^1, T_k^2\right), \quad \boldsymbol{X}_0=\boldsymbol{X}_{init}  \label{con1},\\
& \left|T_k^1\right| \leq T_{max}, \quad \left|T_k^2\right| \leq T_{max}  \label{con2},
\end{align}
\end{subequations}
where $H$ is the prediction horizon. $\boldsymbol{X}=\left[\mathbf{v} \enspace \boldsymbol{\eta} \right]$ is the states of the vehicle. $\mathbf{f}(\cdot)$ is the discrete-time vehicle kinematics and dynamics obtained by
discretizing \eqref{Dynamics} and \eqref{Kinematics} with time step $\Delta t$. $J_{s}$ is the energy consumption generated by ASV thrusters and operations within the predicted range, which represents the stage cost.  $J_{t}$ is the terminal cost, determined by the state at the end of the prediction horizon and the target state. 
 $\boldsymbol{X}_{init}$ is ASV initial condition.
 $T_{max}$ is the upper bound of the thruster input. Additionally, the vehicle sequentially visits all the way points by entering the circle of acceptance (COA) of
each way point by
\begin{align}
&\forall P_j \in \mathcal{P}, \exists \boldsymbol X_{k},\text{such 
 that} \nonumber \\
&\sqrt{\left(x_{k}-{x}_{fj}\right)^2+\left(y_{k}-{y}_{fj}\right)^2} \leq r_{\mathrm{COA}},
\nonumber
\end{align}
 where $k \in[0, H]$. $r_{\mathrm{COA}}$ is the radius of the circle of acceptance.

To solve the control problem in \eqref{em}, the authors in \cite{von1993numerical} transformed the question into a finite dimensional nonlinear program which can be solved by standard sequential quadratic programming (SQP) methods. It is employed to optimize the vehicle trajectory globally.   Here, we briefly describe the basic idea there. Initially, the vehicle's input and state trajectories ranging from the initial to the desired final condition, are uniformly discretized with a constant time increment. Following this discretization, the trajectory optimization issue is recast as a nonlinear programming (NLP) problem. In this context, the discretized states and inputs serve as the decision variables for the NLP. The objective function for the NLP is the total propulsion energy, which is computed based on the discretized inputs. The primary constraint is that any two consecutive states, along with their respective inputs, must adhere to the vehicle's dynamics. Finally, the NLP is addressed numerically to determine the optimal sequence of vehicle thrusts.

However, this approach is likely to suffer from robustness
issues or performance degradation
under model uncertainties and environment disturbance \cite{DCweakness}. Moreover, the computation
for solving each NLP in the above case study is intensive which prohibits it from real-time operation for  resource-limited ASV platforms. The optimal maneuver based on the solution from this method is analyzed in  \cite{2022}. Inspired by this work, we identify two distinctive modes as static mode and
dynamic mode. In the static mode, the vehicle has a constant
surge speed with minimal motion in other DOFs, and the yaw power consumption is negligible. In the dynamic
mode, the vehicle has a nonzero yaw rate, and energy is
used for surge and yaw controls. The dynamic mode
happens only when the vehicle needs to change its direction
(e.g., way point switch). Based on the above characteristic, we further analyze and quantify the energy consumption part of terminal cost. The remaining part is determined by the cross-track error and measured in the form of energy.  Thus, we  design the effectively control  under the constraints of energy and accuracy by EMPC method.

We adopt the EMPC for the following reasons \cite{EMPCReason}: 
 At first, the method systematically optimizes propulsion energy expenditure through direct incorporation of nonlinear ship dynamics and thruster saturation limits within the constrained optimization problem, providing theoretical feasibility guarantees through hard constraint enforcement.
 Secondly, the introduction of the terminal energy cost extends the controller's temporal perspective beyond the finite prediction horizon, enabling compensation for long-term energy impacts.
Thirdly, the proposed energy-path coupling mechanism introduces a novel metric for quantifying navigation precision through energy equivalence.

\section{ECONOMIC MODEL PREDICTIVE CONTROLLER DESIGN}
In this section, we propose an online controller design that
addresses the ASV  control problem with comprehensive conditions formulated in
\eqref{em}. Based on the analysis in Section III, we
develop a controller based on EMPC to control the vehicle thrusters. In the following, the energy consumption cost and the track error cost from the end of the predicted horizon to the target state are derived. Finally, we present the overall schematic
of the EMPC and show the tradeoffs in the vehicle control energy.

\subsection{Cost Function Formulation}
To solve the optimization problem \eqref{em}, we design a controller based on the energy consumption of the thrusters as the stage cost under the EMPC framework. 
In addition, the stage cost and terminal cost are formulated to quantify the energy consumption required for destination convergence and sail accuracy.
We first optimize
the  thruster sequences $\left\{T_{k\lvert t}^1\right\}$ and $\left\{T_{k\lvert t}^1\right\}$ within the
prediction horizon by minimizing the stage cost given by
\begin{equation}
J_s=\left(\sum_{i=1}^2 h_p\left(T_{k \mid t}^i\right)+T^{\mathrm{w}}\right). 
\label{js}
\end{equation}
The terminal cost consists of voyage energy consumption and error penalty:
\begin{equation}
J_t=E+Y,
\label{jt}
\end{equation}
where $E$ is the energy consumption cost and  $Y$ is the track error cost.

Optimization problems for cost functions \eqref{js} and \eqref{jt} are  performed subject to \eqref{con1} and \eqref{con2}. 
In this way, the controller finds motion commands that achieve both dynamic feasibility and energy efficiency.

\subsection{Terminal Energy Consumption Cost Formulation}
\begin{figure}[t]
    \centering
    \includegraphics[scale=0.4]{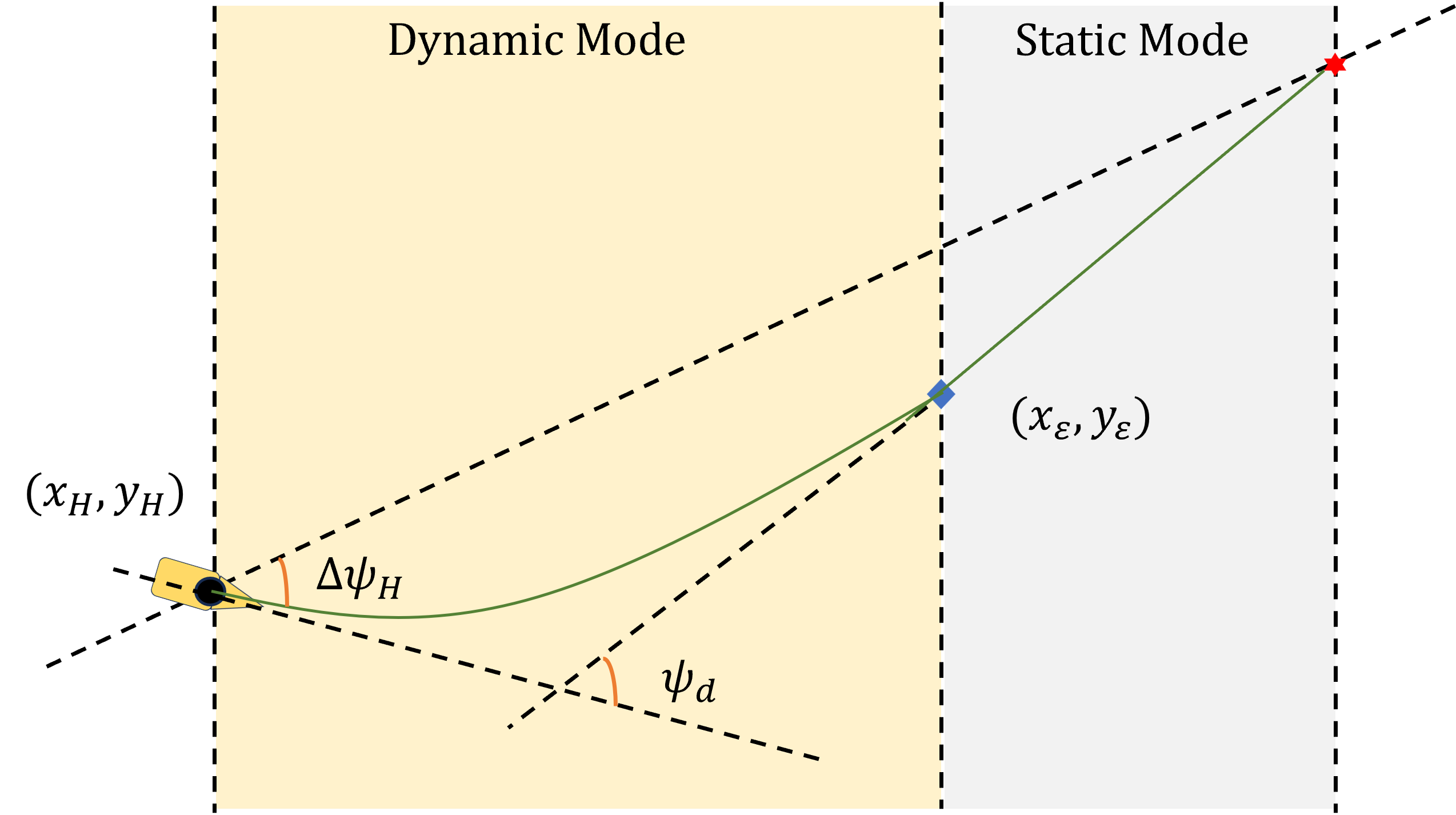}
    \caption{Illustration of the dynamic and static modes  beyond the prediction horizon.}
    \label{ Illustration of the dynamic and static parts.}
\end{figure}

In order to drive the vehicle to the destination, the remaining
energy to reach the destination is supposed to be included
into the optimization problem. We divide the energy into the 
dynamic and static parts and express the terminal cost as
\begin{equation}
E=E_d+E_s,
\nonumber
\end{equation}
where $E_d$ and $E_s$ approximate the dynamic and static parts in
the energy to arrive at the destination, respectively. An illustration of the two components is shown in Fig. \ref{ Illustration of the dynamic and static parts.}. Given that the vehicle control energy is closely
related to its travel time, the travel time in the dynamic
and static parts, denoted by $t_d$ and $t_s$, respectively, are
introduced as extra decision variables in EMPC. To 
 estimate the travel time, we assume that the surge velocity remains unchanged beyond the prediction horizon (i.e., $u_{ k \mid t}=u_{ H \mid t}$ for $k\geq H$). 

\subsubsection{Static Mode Cost}
In the static mode, the thruster power is mainly for surge, whose speed is nearly constant. Thus,
we approximate the total power by
\begin{equation}
P_s=P^{sur}+T^w,
\nonumber
\end{equation}
where $P^{sur} = 2h_p(X_u u_{H \mid t}/2)$ is the power for overcoming
the surge drag force under constant surge speed \cite{2022}.

\subsubsection{Dynamic Mode Cost}
\begin{figure}[t]
    \centering
    \includegraphics[scale=0.5]{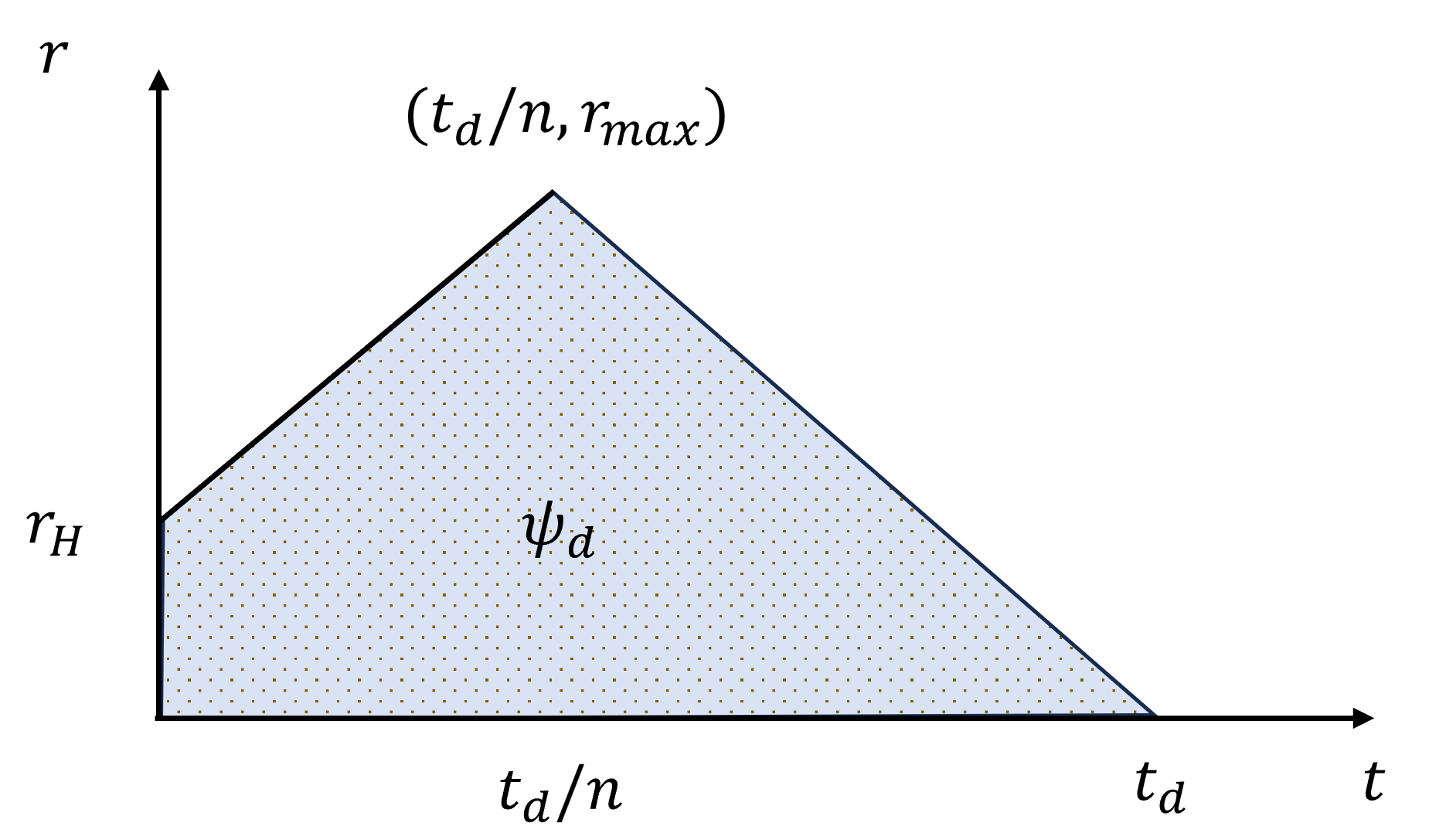}
    \caption{ Approximation of the yaw rate profile during the dynamic stage.}
    \label{ Approximation of the yaw rate profile during the dynamic mode}
\end{figure}

In the dynamic mode, the thruster energy is consumed for surge and yaw controls. The surge power is the same as static stage. 
Denote the variation in the course direction of the vehicle
during the dynamic mode as $\psi_d$ (see Fig. \ref{ Illustration of the dynamic and static parts.}). To achieve the desired heading adjustment, the yaw rate is assumed to follow a characteristic pattern during the dynamic maneuver: increasing to a maximum value $r_{max}$ in the initial phase, then subsequently decreasing to zero in the rest of dynamic phase, as illustrated in Fig. \ref{ Approximation of the yaw rate profile during the dynamic mode}.

Mathematically, the yaw
rate is expressed by 
\begin{flalign}\label{r}
r_{k \mid t}
= \begin{cases}  r_{H \mid t}+a_1\kappa \Delta t, & 0 \leq \kappa \leq \frac{t_d}{n \Delta t}, \\ r_{\max }+a_2\left(\kappa \Delta t-\frac{t_d}{n}\right), & \frac{t_d}{n \Delta t} \leq \kappa \leq \frac{t_d}{\Delta t},\end{cases}
\end{flalign}
where $\kappa = k - H $ is the number of time steps in the dynamic phase after predicting the horizon. 
$\frac{1}{n}$ represents the proportion of the first stage and $r_{max} = 2\psi_d/t_d-r_{H \mid t}/n$. $a_1 = n(r_{max} - r_{H\mid t})/t_d$ represents the $\dot r$ in the first part of the dynamic mode, $a_2 = - n r_{max}/((n-1)t_d)$ represents the $\dot r$ in the rest part. 

From \eqref{r} we derive the power in the dynamic stage following the trapezoid rule as
\begin{equation}
\begin{aligned}
P_d&=P^{sur}+P^{{yaw}}+T^{{w}}\\
&=T^w+\sum_{i=1}^2 \frac{n\left(P_{H \mid t}^i+ P_{H+m^- \mid t}^i\right) }{2n}\\
&+\sum_{i=1}^2 \frac{\left(n-1\right) \left(P_{H+m^+ \mid t}^i+P_{H+n \mid t}^i\right)}{2n},\\
\end{aligned}
\nonumber
\end{equation}
where $P_{H \mid t}^i$, $P_{H+m^- \mid t}^i$, $P_{H+m^+ \mid t}^i$, and $P_{H+n \mid t}^i$ are the power at the
end of the prediction horizon, at the $1/n$-th segment of the dynamic mode when $\dot r = a_1$, at the $1/n$-th segment of the dynamic mode when $\dot r = a_2$, and the end of the dynamic mode,  respectively.

\subsection{Terminal Track Error Cost}
\begin{figure}[t]
    \centering
    \includegraphics[scale=0.5]{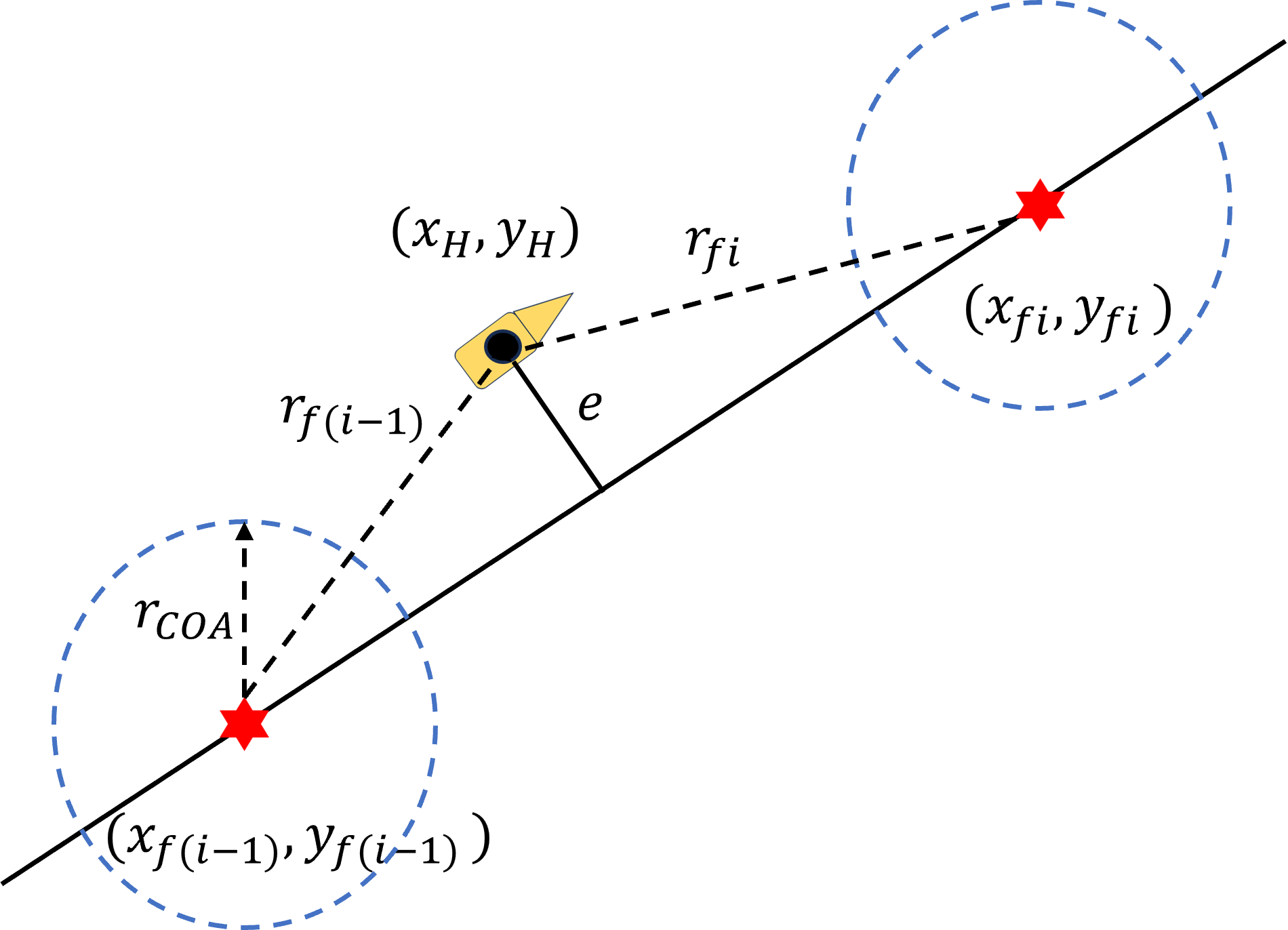}
    \caption{ Schematic of cross-track error.}
    \label{errorcost}
\end{figure}

Due to model uncertainty and environmental disturbances, the ASV inevitably deviates from the expected path. In dynamic marine environments, excessive error can lead to oscillatory behavior or instability. Reducing error is critical for maintaining high path accuracy. Therefore, analysis of the lateral tracking error of unmanned boats is necessary.  Cross-track error is defined as the deviation from the reference path perpendicular to the direction of motion in Fig. \ref{errorcost}.  The cross-track error is imposed as
\begin{equation}
e= \begin{cases}
0,  &r_f \leq r_{COA},\\
\frac{\left|(x_{f(i-1)}-y_{fi})x_{H \mid t}+(y_{f(i-1)}-x_{fi})y_{H \mid t}\right|}{\sqrt{\left(x_{f(i-1)}-x_{fi}\right)^2+\left(y_{f(i-1)}-y_{fi}\right)^2}} ,&r_f \geq r_{COA},
 \end{cases}
 \nonumber
\end{equation}
where $(x_{f(i-1)},y_{f(i-1)})$ is the last way point that vehicle has passed.  $r_f$ is defined as the distance between the vehicle and the nearest way point as 
\begin{equation}
\begin{aligned}    
 r_f = &\min ( \sqrt{\left(u_{H \mid t}-x_{fi}\right)^2+\left(y_m-y_{fi}\right)^2},\\
&\sqrt{\left(x_m-x_{f(i-1)}\right)^2+\left(y_m-y_{f(i-1)}\right)^2}) .
\end{aligned}
 \nonumber
\end{equation}

When the ASV is performing the way point transition inside the COA, the cross-track error is not considered. In order to incorporate the terminal error cost into the EMPC  framework, we approximate the error in the form of energy. 
Within a limited track error, the power to overcome the error comes from surge drag. Therefore, the track error cost is defined as 
\begin{equation}
    Y = \frac{e}{u_{H \mid t}}P^{sur}.
     \nonumber
\end{equation}

In this way, we simplify parameter tuning, reduce oscillations and enhance system stability and robustness.  
This enables the controller to simultaneously consider accuracy and energy efficiency, which leads to a comprehensive improvement in the performance of ASV.

\subsection{ Comprehensive Constrained EMPC}
\begin{figure}
    \centering
    \includegraphics[width=0.9\linewidth]{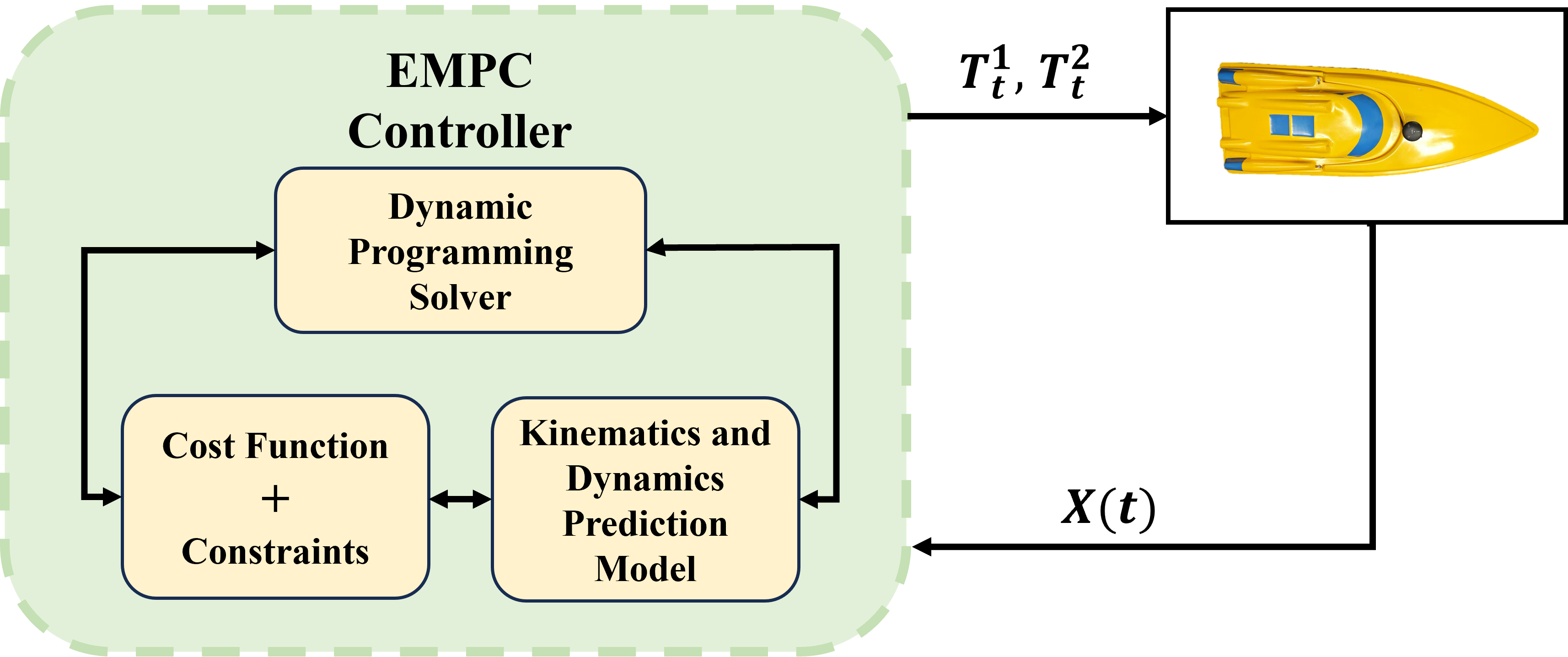}
    \caption{Block diagram for the proposed CC-EMPC}
    \label{schem}
\end{figure}

The block diagram of EMPC with energy and accuracy comprehensive constraints is shown in Fig. \ref{schem}. With the stage cost and estimated terminal cost, the proposed economic model predictive control problem \eqref{em} is further given by 
\begin{subequations}
\begin{align}
& \min _{\left\{T_{k \mid}^1\right\},\left\{T_{k \mid\}}^2\right\}, t_d, t_s} J=\sum_{k=0}^{H-1} L_{k \mid t}+P_d t_d+P_s t_s + Y \label{EMPC}\\
& \text { s.t. } \boldsymbol{X}_{k+1 \mid t}=\mathbf{f}\left(\boldsymbol{X}_{k \mid t}, T_{k \mid t}^1, T_{k \mid t}^2\right), \quad \boldsymbol{X}_{0 \mid t}=\boldsymbol{X}_t ,\\
& \qquad\left|T_{k \mid t}^1\right| \leq T_{max}, \quad\left|T_{k \mid t}^2\right| \leq T_{max}, \quad 0<u_{r, k+1 \mid t}, \\
& \qquad u_{H \mid t }\left(\frac{t_d \sin \left( \Delta \psi_{H \mid t}\right)}{\Delta \psi_{H \mid t}}+t_s\right)=d ,\label{cons}
\end{align}
\end{subequations}
where $\Delta \psi_{H \mid t} = atan2(y_f-y_{H \mid t},x_f-x_{H \mid t})-tan^{-1}(v_{H \mid t} / u_{H \mid t})- \psi_{H \mid t}$ is the difference between the
desired and present course directions of the vehicle at the end
of prediction horizon. It encapsulates three critical components:  
waypoint-oriented bearing angle, velocity vector orientation and current heading offset. \eqref{cons} is a constraint on static mode time $t_s$ and dynamic mode time $t_d$ , which ensures that the vehicle satisfies motion model constraints beyond the prediction horizon.
 From \eqref{EMPC}, we see the tradeoffs in the vehicle EMPC problem: since the power for the task is constant, a shorter travel time reduces energy consumption. However, considering that surge power is
proportional to the squared surge speed, a shorter travel
time will lead to larger energy spent for surge control and track error. Furthermore, the state at the end of the predicted horizon is strategically optimized to simultaneously minimize both the energy consumption and the cross-track error along the prescribed trajectory, thereby enhancing overall path following precision and energy efficiency.

\section{SIMULATION AND EXPERIMENT}
To verify the effectiveness of the proposed predictive controller,  comprehensive constrained economic model predictive control (CC-EMPC) is demonstrated through simulations and field experiments in actual lakes. 
\subsection{Simulations and Analysis}
We use 
the model of section III with CasADi toolbox.   
The sampling time $\Delta t$ is 0.1 $s$, and the
prediction horizon is 1.0 $s$, which means $H = 10$. The upper bound for each thruster $T_{max}$ 
is 10 $N$. For a better performance comparison, the specific kinetic parameters are consistent with the model in \cite{2022}. The performance of EMPC is compared with those obtained from (i) nonlinear model predictive control (NMPC) in \cite{NMPC} and (ii) energy optimal economic model predictive control (EO-MPC) in \cite{2022}. NMPC is based on the standard MPC formula design method, and the cost function is related to the vehicle states, expected states and control inputs. 
\begin{table*}[]
\begin{center}
\caption{Performance Comparison}
\renewcommand{\arraystretch}{1.3} 
\begin{tabular}{c|ccc|ccc|ccc}
\Xhline{1pt}
\multirow{2}{*}{\begin{tabular}[c]{@{}c@{}}Disturbance\\ condition\end{tabular}} & \multicolumn{3}{c|}{Energy consumption} & \multicolumn{3}{c|}{Average cross-track error} & \multicolumn{3}{c}{Travel time} \\ \cline{2-10} 
& NMPC         & EO-EMPC     & CC-EMPC    & NMPC         & EO-EMPC        & CC-EMPC        & NMPC    & EO-EMPC   & CC-EMPC   \\ \hline
\#1 & 2328.4 $J$& 331.9 $J$       & 332.0 $J$& 1.686 $m$& 0.531 $m$         & 0.521 $m$& 36.4 $s$& 338.7 $s$    & 345.1 $s$\\
\#2 & 2321.0$J$& 328.5 $J$      & 328.7   $J$    & 1.680 $m$& 0.542 $m$         & 0.520 $m$         & 36.4 $s$& 337.1 $s$    & 343.6 $s$    \\
\#3 & 2310.2$J$& 323.0 $J$      & 323.2 $J$     & 1.672$m$& 0.685 $m$         & 0.664 $m$         & 36.3 $s$& 334.2 $s$    & 342.7 $s$    \\
\#4 & 2300.0 $J$& 330.1 $J$      & 330.3 $J$     & 1.623 $m$& 0.507 $m$         & 0.477 $m$         & 36.9 $s$& 336.2 $s$    & 343.0 $s$    \\
\#5 & 2323.2 $J$& 342.4 $J$      & 342.6 $J$     & 1.655 $m$& 0.376 $m$         & 0.369 $m$         & 36.4 $s$& 338.6 $s$& 343.4 $s$\\ \Xhline{1pt}
\end{tabular}
\end{center}
\end{table*}
We select the following five conditions with
different forms and intensities of disturbance to make the problem computationally manageable. Among them, the disturbance conditions \#3 – \#5 are selected among the sea surface wind stress data obtained from Global Ocean Data Assimilation System (GODAS) \cite{zhou2022hybrid}. 

1) $\boldsymbol{\tau}_d^u = 0\ N$ and  $\boldsymbol{\tau}_d^v = 0\ N$.

2) $\boldsymbol{\tau}_d^u = 0.015\ N$ and  $\boldsymbol{\tau}_d^v = 0.015\ N$.

3) $\boldsymbol{\tau}_d^x = --0.0003\ N$ and  $\boldsymbol{\tau}_d^y = 0.0799\ N$.

4) $\boldsymbol{\tau}_d^x =-0.0987\ N$ and  $\boldsymbol{\tau}_d^y = 0.0868\ N$.

\begin{figure}
  \centering
  \subfigure[$\boldsymbol{\tau}_d^x$
  ]{\includegraphics[width=0.48\linewidth]{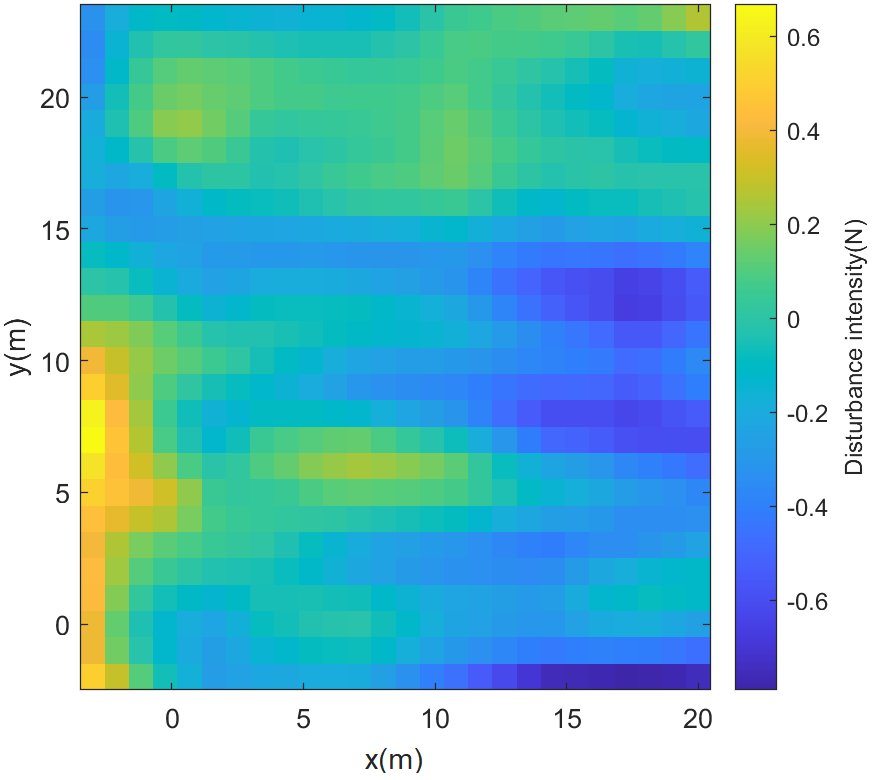}}
  \subfigure[$\boldsymbol{\tau}_d^y$]
  {\includegraphics[width=0.48\linewidth]{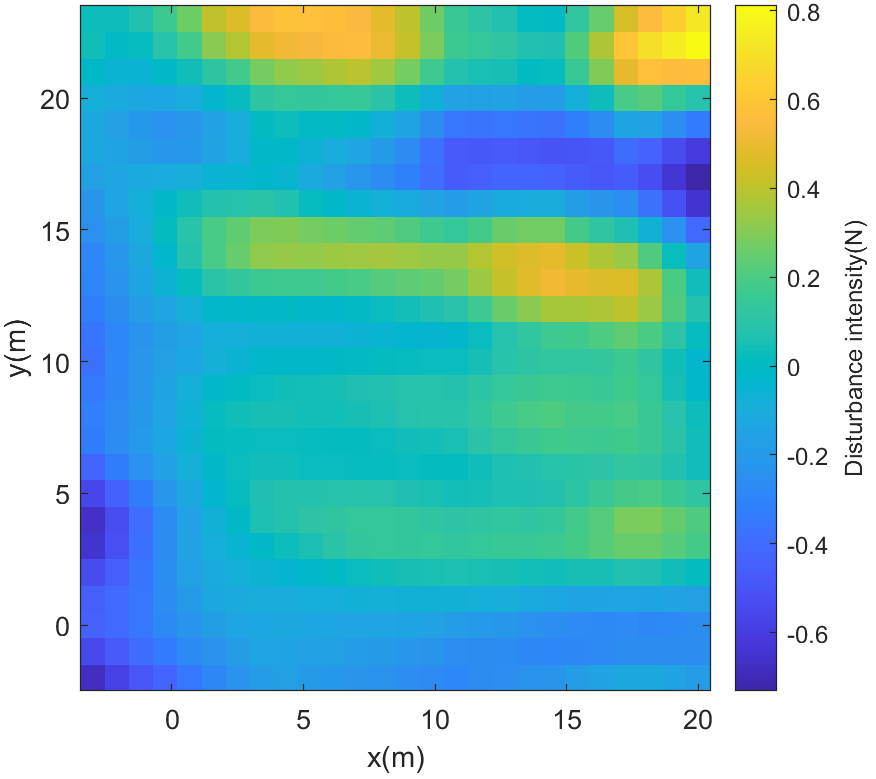}}
  \caption{ Condition \#5 disturbance components distributed with spatial grid. }
  \label{dis}
\end{figure}

5) $\boldsymbol{\tau}_d^x$ and  $\boldsymbol{\tau}_d^y$ are be simulated between spatial grid points as Fig. \ref{dis}.

\begin{figure}[t]
    \centering
    \includegraphics[width=0.95\linewidth]{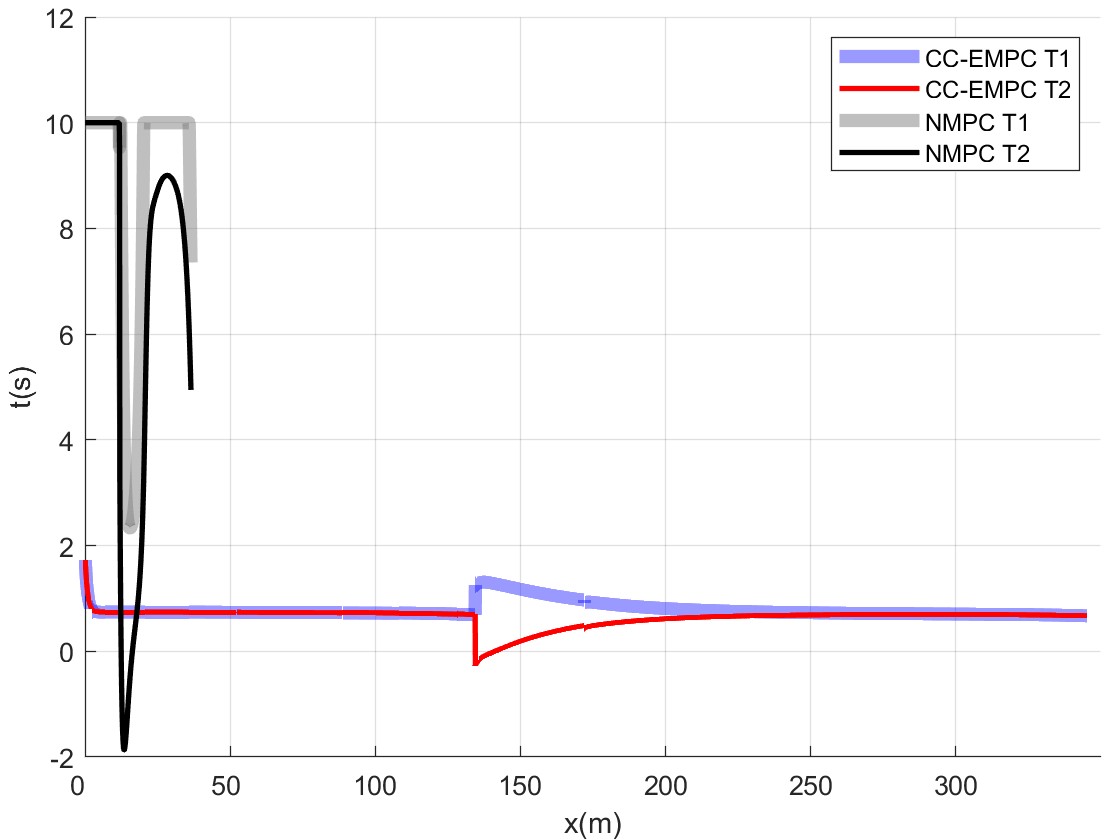}
    \caption{Thruster output of CC-EMPC and NMPC under condition \#1}
    \label{tao}
\end{figure}

\begin{figure*}[tbp]
\centering
\subfigure[]{
\includegraphics[width=0.25\linewidth]{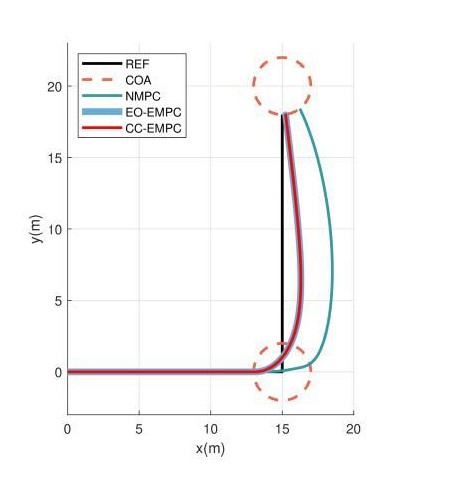} }\hspace{-0mm}
\subfigure[]{
\includegraphics[width=0.25\linewidth]{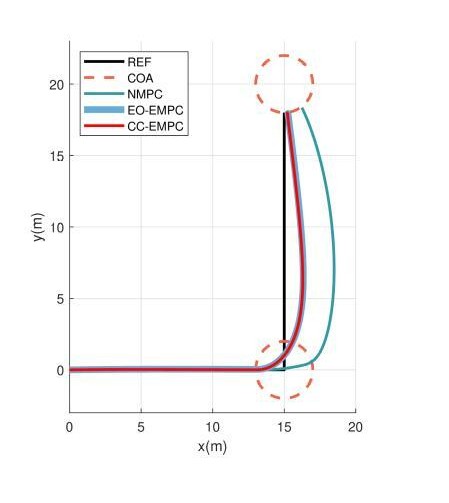} }\hspace{-0mm}
\subfigure[]{
\includegraphics[width=0.25\linewidth]{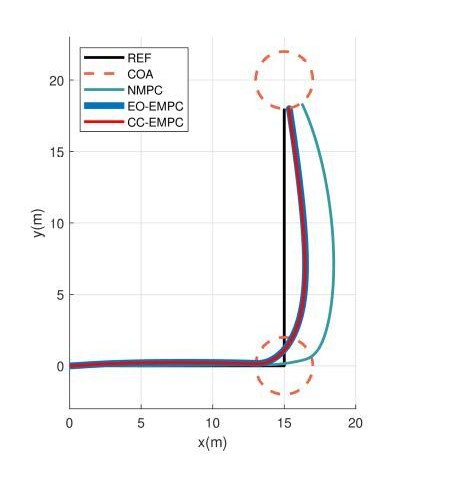} }
\\
\subfigure[]{
\includegraphics[width=0.25\linewidth]{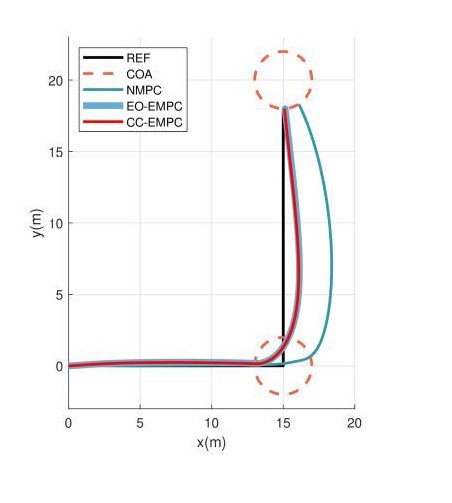} }\hspace{-0mm}
\subfigure[]{
\includegraphics[width=0.25\linewidth]{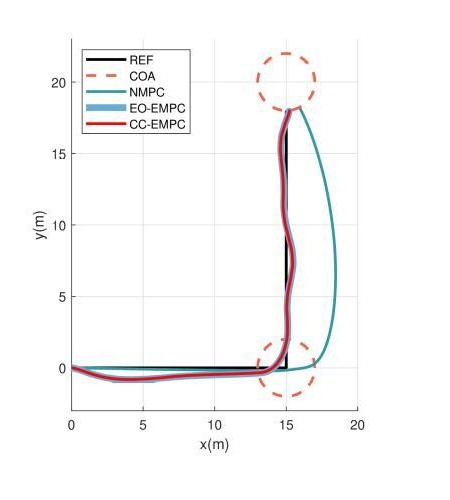} }
\caption{Vehicle trajectories from the NMPC,  EO-EMPC and CC-EMPC under different
disturbance conditions. (a)disturbance condition\#1. (b)disturbance condition\#2. (c)disturbance condition\#3. (d)disturbance condition\#4. (e)disturbance condition\#5. }
\label{tra}
\end{figure*}

Based on the above parameters, we simulate the NMPC, EO-EMPC and the CC-EMPC approach, and summarize the vehicle performance in Table I. The average cross-track error is computed by averaging vehicle deviations from the line between the present and past waypoints when the vehicle is outside the COA.  From Table I and Fig. \ref{tao}, we see that the method based on EMPC has significant improvements in energy consumption and navigation accuracy compared to the traditional MPC method, but at the cost of longer travel time across all the cases. The NMPC controller fully uses the onboard thrust capability in order to generate the fastest possible convergence while respecting the physical limit of thrusters, while the EMPC only aims at minimizing the economic cost. The trajectories resulted from NMPC, EO-EMPC, and CC-MPC are shown in Fig. \ref{tra}. As demonstrated in Fig. \ref{tra}, all control algorithms are able to drive the vehicle to the destination.

Furthermore, compared with EO-EMPC, CC-EMPC incorporates the cost of track error into its cost function, resulting in slightly higher travel energy consumption (0.03\% in case \#1) but a lower cross-track error (a 1.88\% reduction in case \#1). With a fixed disturbance, this difference is more significant. In case \#2, CC-EMPC energy increase 0.06\% but the cross-track error reduces 4.06\%. Under the condition of real perturbed data in case \#3 and \#4, the difference is also  obvious. When the disturbance follows the spatial grid distribution at a certain boundary, this trend also shows some improvement.
In case \#5, the trajectory optimization based on EMPC method is more obvious, while the difference under NMPC method is not significant. The reason is that  EMPC can leverage online optimization to schedule an appropriate control gain to well compensate the disturbances, which is crucial for marine control systems. More specifically, compared with EO-EMPC, CC-EMPC increases energy consumption by 0.06\%, but reduces the cross-track error by 18.61\%, because EO-EMPC solely aims to minimize vehicle energy consumption. \cite{2022} has verified that the energy efficiency of the EO-EMPC method is 10\% lower than that of the optimal solution calculated by the offline direct collocation method. Thus, the performance of CC-EMPC is also very close to that of the optimal solution in terms of energy consumption, while achieving improvements in tracking accuracy and robustness.

\subsection{Experiments in Lake Environment}
\begin{figure}[t]
    \centering
    \includegraphics[width=0.95\linewidth]{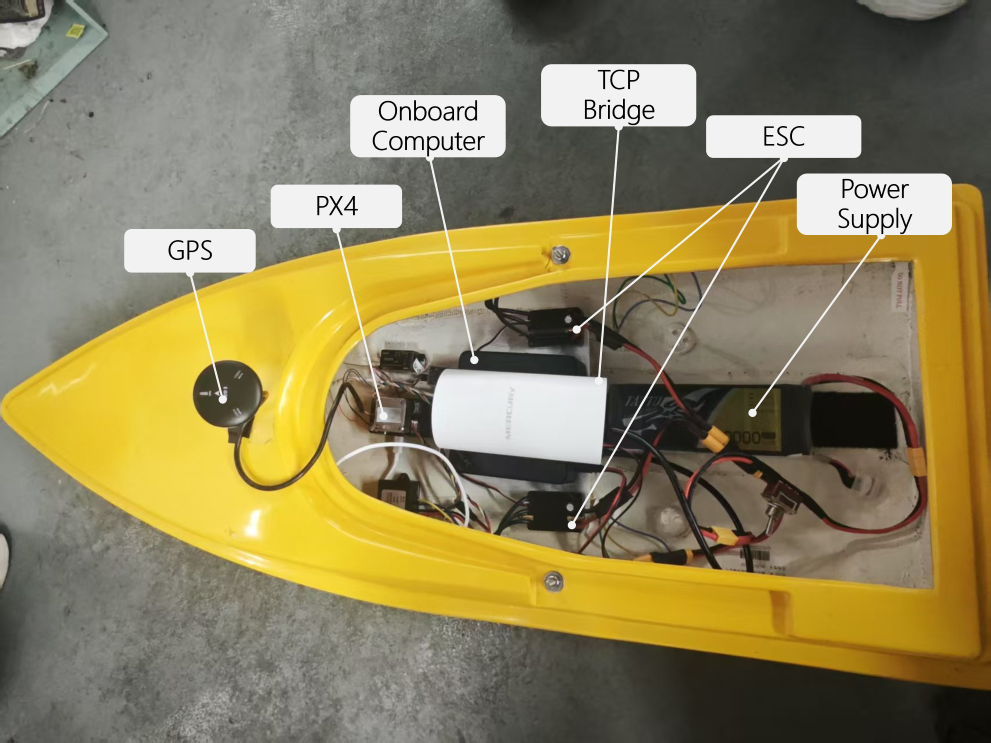}
    \caption{Schematic of the ASV architecture.}
    \label{ASV}
\end{figure}

We present the development of an underactuated  ASV featuring a dual-propeller configuration. Detailed specifications are provided in Table II, while Fig. \ref{ASV}. illustrates the integrated control system architecture, which incorporates multiple critical subsystems to grantee the operational efficiency and system reliability.
 The ground station achieves remote control of the onboard computer through TCP bridge. The onboard computer, running on Ubuntu 22.04 and utilizing Robot Operating
System 2 (ROS2) Humble with Intel N100 CPU, handles various functionalities. Communication between the PX4 flight controller and the onboard ROS environment is facilitated by the uXRCE-DDS (Micro XRCE-DDS) protocol. This communication setup includes nodes and topics that manage specific tasks and data exchange, ensuring robust and efficient coordination.  The flight controller, operating on PX4 version 1.15.2 and configured for the Rover UGA platform, ensures precise navigation and control. Additionally, the ASV is equipped with a GPS system to provide accurate positioning and navigation data. Propulsion and speed regulation are managed by the motor and electronic speed controller (ESC). This comprehensive architecture combines software and hardware elements to support the autonomous and remote operation of the ASV.

\begin{table}[]
\caption{The parameters of ASV}
\begin{tabular}{c c c}
\Xhline{1pt}
Parameter                                     & Nomenclature & Value                  \\ \hline
Mass                                          & $m$            & 7.65 $kg$                \\
Thruster arm & $l$            & 0.1025 m               \\
\multirow{3}{*}{Symmetric mass matrix}        & $m_{11}$          & 12.84 $kg$             \\
                                              & $m_{22}$          & 10.65 $kg$              \\
                                              & $m_{33}$          & 1.86 $kg \cdot m ^2/rad$             \\
\multirow{3}{*}{Drag matrix}                  & $X_u$           & 33.57 $kg/s$           \\
                                              & $Y_v$           & 50.78 $kg/s$           \\
                                              & $N_r$           & 0.31 $kg\cdot m^2/(s\cdot rad)$ \\
Maximum thrust                                & $T_{max}$       & 75 $N$                    \\
Thrust dead zone boundary         & $T_{min}$         & 10 $N$                    \\ 
Control frequency & $f$ & 10 $Hz$
\\
\Xhline{1pt}
\end{tabular}
\end{table}

\begin{figure}[t]
    \centering
    \includegraphics[width=0.95\linewidth]{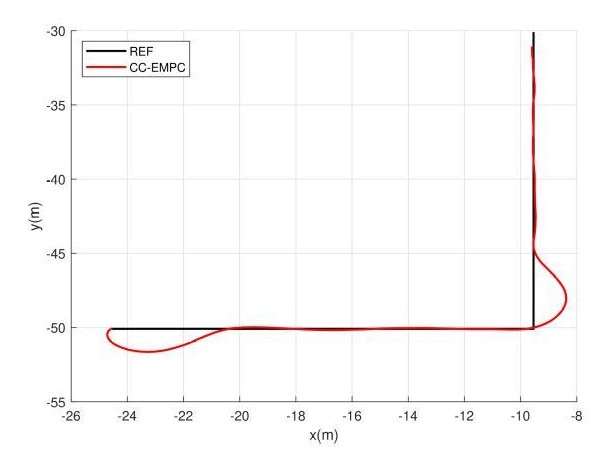}
    \caption{Path tracking result of CC-EMPC
controller in real lake}
    \label{real}
\end{figure}

\begin{figure}[t]
    \centering
    \includegraphics[width=0.95\linewidth]{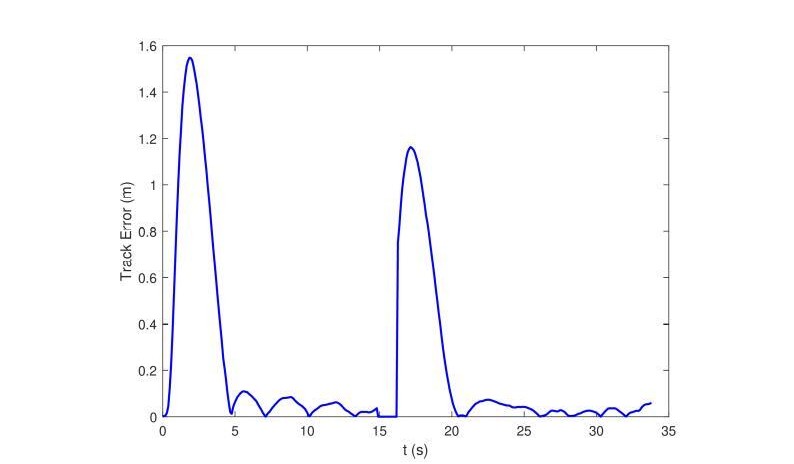}
    \caption{Cross-track error of ASV in real lake}
    \label{real_error}
\end{figure}

Fig. \ref{real} displays the tracking performance of unmanned boats using the CC-EMPC method in real lake environments. Under real disturbance conditions and  sensor noise, our proposed method follows path tracking tasks well on low performance CPU. When the ship's initial heading angle deviates from the course, it quickly adjusts the direction and returns to the normal course, as shown in Fig. \ref{real_error}. The observed performance enhancement primarily originates from systematic integration of terminal tracking error penalty terms in the CC-EMPC cost function. This design feature proves particularly effective in straight-line path following scenarios, where the control architecture demonstrates active compensation capabilities against persistent unknown disturbances (e.g., wind/wave coupling effects).   After switching way point, it performs turns well and continue to track. 
At a speed close to 1 m/s and a control frequency of 10 Hz, the average cross-track error of the ASV is 0.22 m.
The main sources of the error are the adjustment of the ship's initial heading angle and the turning process.
The experimental results show that the proposed method is feasible to deploy on ASV in real environment.

\section{CONCLUSIONS}
In this paper, we present a novel EMPC algorithm
for the path following control of an ASV. To achieve energy efficiency close to that of the optimal solution, the terminal cost in EMPC is formulated as the approximate energy-to-go and the energy required for track-error correction. The energy required to reach the waypoint is partitioned into dynamic and static components. These components are then parameterized based on their durations, destination, and vehicle dynamics, following the method in \cite{2022}.  We further optimized the cost function by converting the error of the vehicle at the end of the prediction horizon into the form of energy, thereby improving the accuracy of path following. The simulation results of tracking under disturbance data from a real marine environment demonstrate the advantages of our proposed CC-EMPC path-following control. While the energy consumption approaches that of the optimal solution, the method reduces the cross-track error. Both simulation and experimental results verify the effectiveness and feasibility of this method. Future work will
 focus on (i) the recursive feasibility and closed-loop stability of the EMPC
control, (ii) validation on more types of tracking tasks and disturbance conditions,
and (iii) extension to obstacle avoidance and coordination control problems.

\section*{ACKNOWLEDGMENTS}
Thanks for the data service provided by the Oceanographic Data Center, Chinese Academy of Sciences (CASODC).

\bibliographystyle{IEEEtran}
\bibliography{reference}

\end{document}